# Nonreciprocal Magnon Hanle Effect in Antiferromagnetic α-$Fe_2O_3$

Janine Gückelhorn[1,2], Sebastián de-la-Peña[3], Monika Scheufele[1,2], Matthias Grammer[1,2], Matthias Opel[1], Stephan Geprägs[1], Juan Carlos Cuevas[3], Rudolf Gross[1,2,4], Hans Huebl[1,2,4], Akashdeep Kamra[3], and Matthias Althammer[1,2]

[1]Walther-Meissner-Institut, Bayerische Akademie der Wissenschaften, 85748 Garching, Germany
[2]TUM School of Natural Sciences, Technische Universität München, 85748 Garching, Germany
[3]Condensed Matter Physics Center (IFIMAC) and Departamento de Física Teórica de la Materia Condensada, Universidad Autónoma de Madrid, 28049 Madrid, Spain
[4]Munich Center for Quantum Science and Technology (MCQST), 80799 München, Germany

**The precession of the magnon pseudospin about the equilibrium pseudofield, the latter capturing the nature of magnonic eigen-excitations in an antiferromagnet, gives rise to the magnon Hanle effect. Its realization via electrically injected and detected spin transport in an antiferromagnetic insulator demonstrates its high potential for devices and as a convenient probe for magnon eigenmodes and the underlying spin interactions in the antiferromagnet. Here, we observe a nonreciprocity in the Hanle signal measured in α-$Fe_2O_3$ using two spatially separated Pt electrodes as spin injector and detector. Interchanging their roles alters the detected magnon spin signal. The recorded difference depends on the applied magnetic field and reverses sign when the signal passes its nominal maximum at the so-called compensation field. Our findings unlock the high potential of antiferromagnetic magnonics towards the realization of electronics-inspired phenomena.**

***Index Terms*— Antiferromagnetic Materials, Dzyaloshinskii-Moriya Interaction, Magnonics, Spin Currents**

## I. Introduction

Quantized excitations of the spin system in magnetically ordered materials, i.e. magnons, offer a unique platform for future information technology. In combination with the possibility of electrical injection and detection, magnonic spin transport paves the way for the realization of novel spintronic devices [1,2]. Antiferromagnetic materials host pairs of spin-up and spin-down magnons as their quantized spin excitations. We describe them in terms of a magnonic pseudospin [3,4]. Its close analogy to the electronic spin led to the prediction of novel fascinating magnon transport phenomena [3-6]. We observe and investigate these effects in the electrically insulating antiferromagnetic oxide α-$Fe_2O_3$ (hematite), which harbors a finite Dzyaloshinskii-Moriya interaction (DMI). Together with the easy-plane anisotropy, this gives rise to a slight canting of the antiferromagnetic sublattice magnetizations in the (0001) plane at room temperature and, hence, to a residual net magnetic moment [7,8]. This manifests itself in a coherent precession of the magnonic pseudospin about the pseudofield [3]. The combination of α-$Fe_2O_3$ thin films and heavy metal Pt electrodes exhibiting a large spin-orbit coupling represents a prototypical bilayer system to investigate electrical magnon injection and detection [4,5], spin-Hall magnetoresistance [9], as well as antiferromagnetic magnon propagation [4-6,10].

## II. Experimental

We study the diffusive spin transport in a non-local geometry in two-terminal devices, each consisting of two Pt electrodes on epitaxial, electrically insulating α-$Fe_2O_3$ thin films with different thicknesses (Fig. 1(a)). Spin-orbit interaction in Pt causes spin-charge coupling and allows for electrical injection and detection of magnon spin currents. Using a current reversal method enables us to discriminate between effects of electrical and thermal origin, i.e. to eliminate signals due to Joule heating. To characterize the sample we perform angle-dependent electrical transport measurements while rotating the orientation φ of an external magnetic field **H** within the thin film plane (Fig. 1(a)) [10]. A dc charge current with the magnitude of $I_{\mathrm{inj}} = 500\ \mu\mathrm{A}$ is applied first to the left Pt electrode leading to spin injection via the spin Hall effect (SHE) into the α-$Fe_2O_3$ film. The resulting magnon accumulation diffuses along the film plane, representing a diffusive current of magnonic pseudospin. For "forward propagation", the magnon density can be detected electrically via the inverse SHE as a voltage signal $V_{\mathrm{det}}$ at the right Pt electrode in a distance $d$ (Fig. 1(a)). In the following, we consider the electrical magnon spin signal defined by $R^{\mathrm{el}} = V_{\mathrm{det}}/I_{\mathrm{inj}}$. The process is most effective for φ = 90° and 270°.

In our experiments, we demonstrate the electrical magnon injection, diffusive magnon transport, and magnon detection [4,5]. We observe the coherent precession of the magnonic pseudospin caused by the easy-plane anisotropy and the Dzyaloshinskii-Moriya interaction in α-$Fe_2O_3$ [4,5]. Varying the magnitude of the external magnetic field allows us to control the precession frequency and, hereby, to interpret our observation as the magnonic analogue of the electronic Hanle effect [4,5]. We detect a maximum of the observed magnon spin signal not in zero field, but at a finite value of 6 T where the effects of canting and easy-plane anisotropy compensate, resulting in zero magnon pseudofield (black arrow in Fig. 1(b)). At this compensation field, the magnon pseudospin precession is suppressed and the magnon spin signal becomes largest [4]. It decreases for smaller (green arrow) and higher magnetic fields (orange arrow).

In thick films, we additionally observe an oscillating behavior of the magnon spin signal in the high magnetic field range as well as an offset signal in the low magnetic field regime (not shown here) [5]. We attribute this offset signal to the presence of finite-spin low-energy magnons [5].

In a second step, we now interchange the injector and detector electrodes, i.e. $I_{inj}$ is injected at the right Pt electrode and the voltage $V_{det}$ is detected at the left Pt strip in the very same sample ("backward propagation") [10]. The measured magnon spin signal $R^{el}$ is plotted in Figure 1(c) versus the angle φ of the in-plane magnetic field for three different magnitudes $\mu_0 H$. The full circles correspond to the forward propagation direction, while open circles represent the backward propagation direction. Evidently, all curves appear to exhibit the $\sin^2 \varphi$ angular dependence, characteristic of a factor $\sin \varphi$ contributed by both of the injection and detection processes. However, a careful examination shows that there are differences in the magnon spin signal between the two propagation directions for $\mu_0 H = 5$ T and 7 T, predominantly at $\varphi = 90°$ and 270°, where $R^{el}$ is largest. This represents evidence for a nonreciprocal diffusive magnon propagation in α-$Fe_2O_3$ [10].

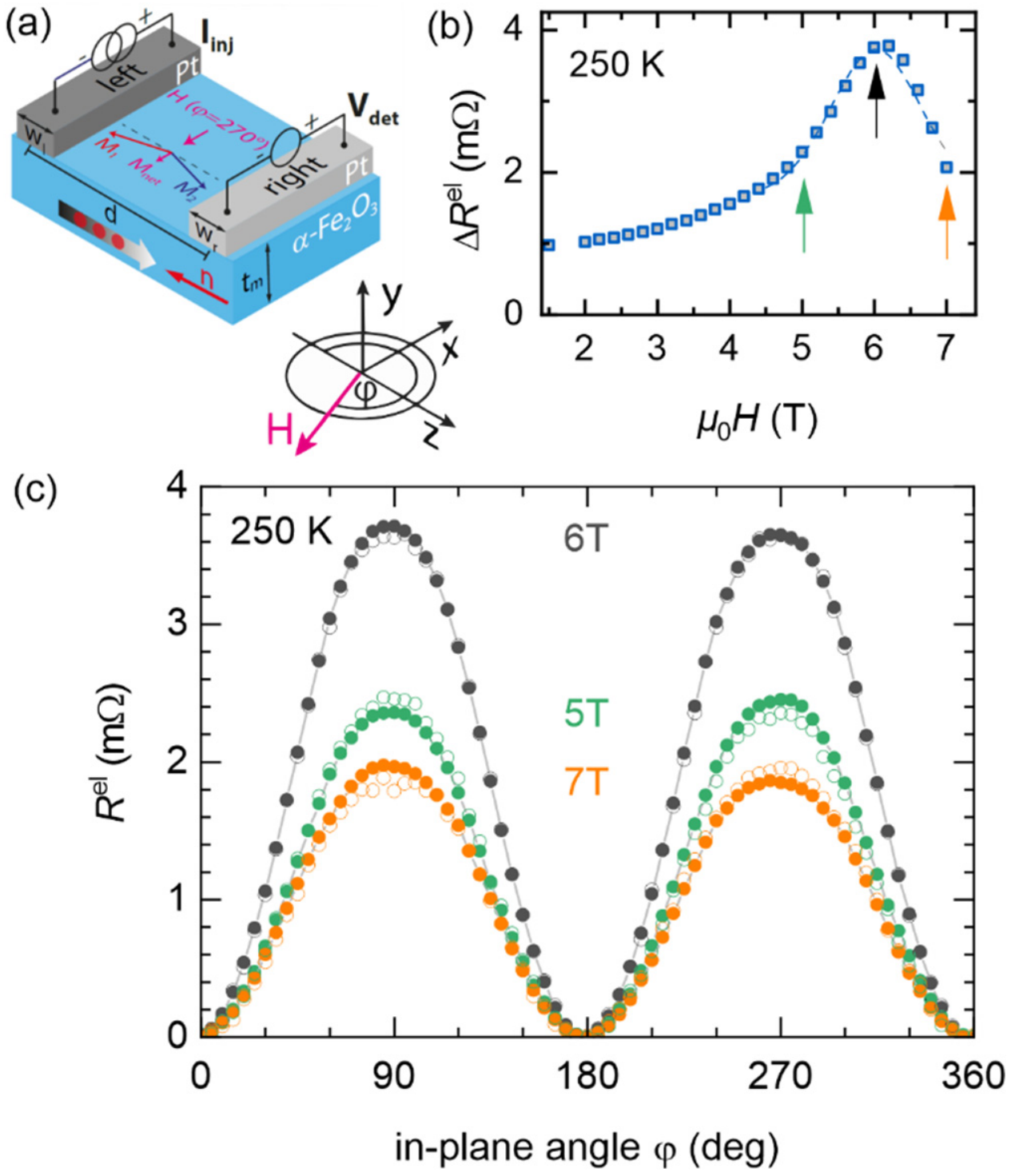

Fig. 1. (a) Sketch of the sample configuration for the forward magnon propagation direction, the electrical wiring scheme, and the in-plane rotation angle φ of the applied magnetic field $\mu_0\mathbf{H}$. The corresponding net magnetization $\mathbf{M}_{net}$ is aligned along the applied magnetic field $\mu_0\mathbf{H}$ ($\mathbf{M}_{net} \parallel \mathbf{H}$), while the Néel vector $\mathbf{n} \perp \mathbf{H}$. The α-$Fe_2O_3$ film (blue) is $t = 89$ nm thin, the center-to-center distance of the Pt electrodes (grey) measures $d = 1.2$ µm. (b) Magnetic field dependence of the amplitude of the magnon spin signal $R^{el}$ at the detector electrode, recorded at $T = 250$ K. (c) Angle dependence of the magnon spin signal $R^{el}$ measured at different magnetic field magnitudes. The full and open circles depict the measured signal in the forward and backward propagation configurations, respectively. A constant offset arising from the experimental setup has been subtracted from the data.

We deduce that our experimentally observed nonreciprocity originates from an antisymmetric magnonic pseudofield and find it to be angle dependent as $\sin \varphi$. This angle dependence is reminiscent of a related, but distinct, nonreciprocity of the magnon dispersion found in $Y_3Fe_5O_{12}$/$Gd_3Ga_5O_{12}$ heterostructures [11], attributed to an interfacial DMI. To examine this potential origin for our case, atomistic spin modeling of α-$Fe_2O_3$ taking into account its exact crystal structure is desirable and will hopefully be motivated by our findings.

## III. Conclusion

We observe diffusive magnon transport in electrically insulating α-$Fe_2O_3$ thin films with a precession of the magnon pseudospin. The spin transport direction-dependent pseudofield leads to a nonreciprocity of the effect, which is found to be controllable via the applied magnetic field [10]. The observed nonreciprocal response in the readily available α-$Fe_2O_3$ films opens interesting opportunities for realizing exotic physics predicted so far only for antiferromagnets with particular crystallographic structures.

We gratefully acknowledge financial support from the German Research Foundation under Germany's Excellence Strategy – EXC-2111 – 390814868 and project AL2110/2-1 and the Spanish Ministry for Science and Innovation – AEI Grant CEX2018-000805-M.